\documentstyle[12pt,epsf]{article}
\newcommand{\beq}{\begin{equation}}
\newcommand{\eeq}{\end{equation}}
\newcommand{\beqa}{\begin{eqnarray}}
\newcommand{\eeqa}{\end{eqnarray}}
\newcommand{\boldtau}{\mbox{\boldmath $\tau$}}

\newcommand{\boldpi}{\mbox{\boldmath $\pi$}}
\newcommand{\slashT}{\slash \hspace{-0.4em}T}

\begin{document}

\begin{titlepage}

\vspace{2.5cm}

\begin{center}
{\Large\bf The Electric Dipole Form Factor of the Nucleon}

\vspace{2.0cm}

{\bf W.H. Hockings}\footnote{{\tt hockings@physics.arizona.edu}}
and 
{\bf U. van Kolck}\footnote{{\tt vankolck@physics.arizona.edu}}

\vspace{0.8cm}
{\it
 Department of Physics\\ 
 University of Arizona\\
 Tucson, AZ 85721}
\end{center}

\vspace{1.5cm}

\begin{abstract}
The electric dipole form factor of the nucleon stemming from
the QCD $\bar{\theta}$ term is calculated in
chiral perturbation theory in leading order.
To this order, the form factor originates from
the pion cloud.
Its momentum dependence is proportional to a
non-derivative time-reversal-violating pion-nucleon coupling,
and the scale for momentum variation 
---appearing, in particular, in the radius of the form factor---
is the pion mass. 
\end{abstract}

\vspace{2cm}
\vfill
\end{titlepage}

\setcounter{page}{1}

Electric dipole moments (EDMs) of the neutron and of atoms
have long been of interest
because precise experiments are possible,
providing strict bounds on possible 
sources of $CP$ violation. (For a review of both experimental
and theoretical results, see for example Ref. \cite{KLbook}.)
New experiments with ultracold neutrons in preparation
at various laboratories (LANL, PSI, ILL, M\"unchen)
promise to provide even higher precision for the neutron
EDM than the current value $d_n=(-1.0\pm 3.6) \cdot 10^{-26} \, e$ cm 
\cite{currentbound}.
While neutron experiments yield bounds on the neutron EDM
directly, atomic effects are sometimes sensitive to the nuclear
Schiff moment, which in turn receives a contribution from the 
radius of the nucleon electric dipole form factor (EDFF).

Nucleon properties are notoriously difficult 
to calculate directly from QCD.
At low momenta, 
$Q\sim m_\pi \ll M_{QCD}$, where $m_\pi$ is the pion mass
and $M_{QCD}\sim 1$ GeV is the typical
mass scale in QCD, they can nevertheless be described
in terms of an effective field theory (EFT) involving
nucleons and pions (and delta isobars), known
as chiral perturbation theory ($\chi$PT).
(For an introduction, see for example Refs. \cite{Wbook,ulfreview}.)
In this theory, 
long-range effects due to the light pions are
separated from short-range effects due to all higher-energy
degrees of freedom. 
Observables are systematically expanded in powers of $Q/M_{QCD}$
(times functions of $Q/m_\pi$).
The pion cloud around the nucleon can generate
contributions that are non-analytic in momenta and
quark masses, and are calculable in terms of pion and
nucleon (and delta) parameters. 
In contrast, short-distance physics contributes powers of $Q^2$ and
$m_\pi^2$ that can only be calculated using detailed
knowledge of the QCD dynamics.

The leading non-analytic contribution to the nucleon EDM 
coming from the QCD $\bar\theta$ angle was first calculated 
in $\chi$PT in Ref. \cite{CDVW79}, and rederived many times since,
see {\it e.g.} Ref. \cite{B00}.
A measurement of the neutron EDM cannot disentangle 
this long-range effect from short-range, analytic terms. 
Since a cancellation between different functions
of  $m_\pi^2$ is unlikely, the leading non-analytic contribution 
serves as an estimate of the EDM.

Here we consider the full nucleon EDFF in leading order in $\chi$PT.
The other nucleon form factors have all been calculated previously
in this theory
at the first few orders:
electric \cite{EMFF}, magnetic \cite{EMFF}, and anapole \cite{AFF}.
Needless to say, the momentum dependence of
the EDFF is not as easily accessible experimentally. 
However, it might become of interest after 
the neutron EDM (the EDFF at $Q^2=0$) is measured.
As we will show below, the EDFF is in leading order 
a calculable isovector function of $Q^2/m_\pi^2$, entirely
determined by the pion cloud.
It is $m_\pi^2$ that determines the scale of variation in
the form factor ---an example of a low-energy theorem.
In particular, the radius of the EDFF is fixed by $m_\pi^2$,
as first derived 
in Ref. \cite{scott}.
If measured, the neutron EDFF  would allow a 
determination of the $T$-violating pion-nucleon coupling
(within the precision of the $\chi$PT expansion). 

At $Q\sim m_{\pi}$, the nucleon is essentially
non-relativistic, because
$m_N\sim M_{QCD}$. 
Pions must explicitly be accounted for in the theory,
since they are the light (pseudo-)Goldstone bosons 
corresponding to the spontaneous breaking of (explicitly-broken) 
chiral $SU_L(2)\times SU_R(2)\sim SO(4)$ symmetry down to 
$SU_{L+R}(2)\sim SO(3)$.  
Also, since the mass difference between the $\Delta$ isobar and the nucleon
is comparable to the pion mass, 
$m_\Delta -m_N\sim 2m_{\pi}$, the $\Delta$ should be included as well 
(which is not much of a problem since at these momenta, 
the $\Delta$ is also a nonrelativistic object).  
This can be done by first constructing the most general Lagrangian involving 
nucleons, pions, and deltas that transforms under the symmetries of QCD in the 
same way as QCD itself.  Along with this, one needs a power-counting scheme 
so that interactions can be ordered according to the expected size of their 
contributions.  
The strong-interaction
Lagrangian contains an infinite number of terms that can be grouped 
according to the index $\Delta$:
\begin{equation}
 {\cal L}=\sum_{\Delta=0}^{\infty}{\cal L}^{(\Delta)}
          \quad,\quad\Delta\equiv d+f/2-2,
\end{equation}
\noindent 
where $d$ is the number of derivatives, powers of $m_{\pi}$ and/or powers of 
$m_\Delta -m_N$, and $f$ is the number of fermion fields.
Electromagnetic interactions are proportional to the small charge $e$,
and it is convenient to account for factors of $e$
by enlarging the definition of $d$ accordingly.

The technology for constructing the Lagrangian is well known,
see for example Ref. \cite{Wbook}.
We will need the following $T$-conserving terms,
which either obey chiral symmetry or break it in the same way
as the quark-mass terms do:
\begin{equation}
{\cal L}^{(0)}_{str/em}=
\frac{1}{2}D_{\mu}\boldpi \cdot D^{\mu}\boldpi
-\frac{1}{2D}m_{\pi}^{2}\boldpi^{2}
+\bar{N}iv\cdot D N
-\frac{g_{A}}{f_{\pi}} \bar{N}\big(S\cdot D\boldtau\cdot\boldpi\big)N
+ \ldots
\label{Lstr0}
\end{equation}
\noindent 
Here $\boldpi$ denotes the pion field in a 
stereographic projection of $SO(4)/SO(3)$, with 
$D=1+\boldpi^2/4f_\pi^2$ and  $f_{\pi}=93$ MeV 
the pion decay constant, 
$N=(p \; n)^T$ is a heavy-nucleon field of 
velocity $v^\mu$ and spin $S^\mu$
($S^\mu=(0, \vec{\sigma}/2$) in the nucleon rest frame
where $v^\mu=(1, \vec{0}$)),
$D_{\mu}$ is the covariant derivative,
with 
$(D_{\mu}^{(\pi)})_{ab}
=D^{-1}(\delta_{ab}\partial_\mu + ie\epsilon_{3ab}A_\mu)$ 
for a pion and
$D_{\mu}^{(N)}=\partial_\mu 
+ i\boldtau\cdot(\boldpi\times D_{\mu}\boldpi)/4f_\pi^2 
-ieA_\mu(1+\tau_{3})/2$ 
for a nucleon, 
and ``$\ldots$'' stands 
for other interactions with more pions, nucleons and/or 
delta fields that are not explicitly needed in the following.  
Note that the pion mass term comes from
the explicit breaking of chiral symmetry by the 
average quark mass $\bar{m}=(m_u+m_d)/2$,
so $m_\pi^2={\cal O}(M_{QCD} \bar{m})$.
Isospin breaking from the quark mass difference,
which is proportional to 
$\bar{m} \varepsilon ={\cal O}(m_\pi^2\varepsilon/M_{QCD})$ 
with $\varepsilon=(m_d-m_u)/(m_u+m_d)\simeq 1/3$,
does not appear at this order.
Note also that at this order the nucleon is static and couples
only to longitudinal photons. Kinetic corrections and magnetic couplings 
have relative size ${\cal O}(Q/M_{QCD})$ 
and appear in ${\cal L}^{(1)}_{str/em}$.
The same is true for the delta isobar, including
the nucleon-delta transition through coupling to a
transverse photon.
The pion-nucleon coupling in Eq. (\ref{Lstr0}) 
is not determined from symmetry, 
but is expected to be ${\cal O}(1)$, and indeed $g_{A}=1.267$.
The Goldberger-Treiman relation 
$g_{A}m_{N}=f_{\pi}g_{\pi NN}$ holds in lowest order.
A term in ${\cal L}^{(2)}_{str/em}$ provides an 
${\cal O}((m_\pi/M_{QCD})^2)$ correction
that removes the so-called Goldberger-Treiman discrepancy.

In addition, we need $P$- and $T$-violating interactions. The exact form of 
these interactions will depend on the mechanism of $CP$ violation.
Just above $M_{QCD}$, $T$-violating interactions among
quarks and gluons can be classified
through their scale dimensions, starting with the $\bar{\theta}$ term.
In a basis where the quark fields $q=(u \; d)^T$ have been 
appropriately rotated 
\cite{Baluni}, we can write
\begin{equation}
{\cal L}^{QCD}_{\slashT}=m_\star \bar{\theta} \; \bar{q}i\gamma_{5}q
   + \ldots,
\label{TviolQCD}
\end{equation}
where $m_\star= m_u m_d/(m_u+m_d)=\bar{m}(1-\varepsilon^2/4)/2
\simeq \bar{m}/2$.
Here ``\ldots'' represents higher-dimension operators
---such as the quark EDM and color EDM,
the Weinberg three-gluon operator, and four-quark interactions---
which we will neglect.  

The $\bar\theta$ term generates $T$-violating interactions in the 
low-energy EFT.
It is the fourth component of the $SO(4)$ vector 
$P=(\bar{q} \boldtau q, \bar{q} i\gamma_{5} q)$, 
so it generates EFT interactions that transform as $T$-violating, 
fourth components 
of $SO(4)$ vectors made out of hadronic fields. 
The lowest-dimension operator of this type is
\begin{equation}
{\cal L}^{(1)}_{\slashT}=-\frac{\bar{g}_{0}}{D}
\bar{N}\boldtau\cdot\boldpi N
\label{LTviol-1}
\end{equation} 
\noindent 
where $\bar{g}_{0}$ is an $I=0$ $T$-violating $\pi NN$ coupling.
{}From dimensional analysis, 
$\bar{g}_{0}= {\cal O}(m_\star \bar{\theta}/f_\pi)
={\cal O}(m_\pi^2 \bar{\theta}/f_\pi M_{QCD})$. 
Because of the factor of $m_\pi^2$
implicit in $\bar{g}_{0}$, ${\cal L}^{(1)}_{\slashT}$  
effectively has index $\Delta=1$.

Interactions with higher indices can be constructed in similar ways.
$T$-violating nucleon-delta transitions through pion emission, for example,
involve at least one derivative, and are thus suppressed by 
at least ${\cal O}(Q/M_{QCD})$ relative
to those stemming from the Lagrangian (\ref{LTviol-1}).
Among the higher-order operators, particularly relevant here are
short-range contributions to the nucleon EDM,  
 \begin{equation}
{\cal L}^{(3)}_{\slashT}=\bar{N}(\tilde{d}_{0}+\tilde{d}_{1}\tau_{3})
                           (S_{\mu}v_{\nu}-S_{\nu}v_{\mu})N F^{\mu\nu},
\label{LTviol2}
\end{equation}
\noindent 
where $\tilde{d}_0$ ($\tilde{d}_1$) is a short-range contribution to the
isoscalar (isovector) EDM of the nucleon. 
{}From dimensional analysis,
$\tilde{d}_{i} ={\cal O}(em_\star \bar{\theta}/M_{QCD}^2)
={\cal O}(em_\pi^2 \bar{\theta}/M_{QCD}^3)$.
Direct  short-range contributions to the 
momentum dependence of the EDFF first appear
in ${\cal L}^{(5)}_{\slashT}$,
being further suppressed by ${\cal O}((Q/M_{QCD})^2)$.

Here we denote by $iJ^\mu_{ed}$ the 
$T$-violating nucleon current
that interacts
with the electron current $-ie \, \overline{e} \gamma^\mu e$ via the 
photon propagator $iD_{\mu\nu}=- i(\eta_{\mu\nu}/q^2 +\ldots)$ 
to produce a contribution
\begin{equation}
iT= -i e \, \overline{e}(k') \gamma^\mu e(k) \, D_{\mu\nu}(q) 
        \, \overline{N}(p') J^\nu_{ed}(q) N(p),
\end{equation}
to the electron-nucleon $S$ matrix.
We have $q^2=(p-p')^2\equiv -Q^2<0$.
We can write
\begin{equation}
J^\mu_{ed}(q)= 2 
               \left(F^{(0)}_D(-q^2)
           +F^{(1)}_D(-q^2) \tau_3\right) (S^\mu v\cdot q- S\cdot q v^\mu),
\label{cur}
\end{equation}
where $F^{(0)}_D(Q^2)$ ($F^{(1)}_D(Q^2)$)
is  the isoscalar (isovector) EDFF
of the nucleon, with
$F^{(i)}_D(0)=d_i$ the corresponding EDM.

\begin{figure}[t]
\begin{center}
\epsfxsize=12cm
\centerline{\epsffile{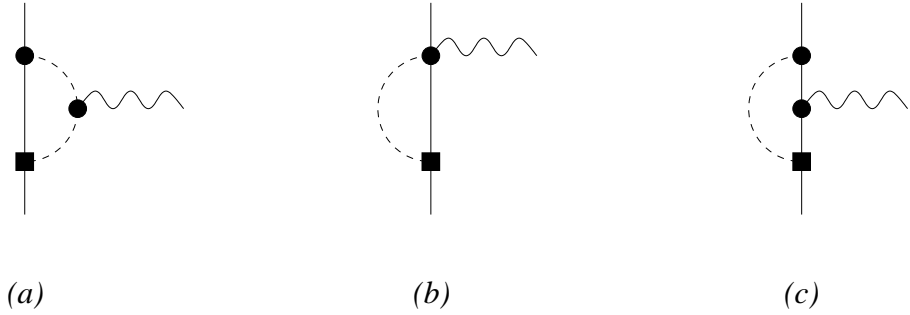}}
\end{center}
\caption{One-loop diagrams contributing to the nucleon 
electric dipole form factor in
leading order. Solid, dashed and wavy lines 
represent nucleon, pion and photon,
respectively; circles and squares stand for interactions
from ${\cal L}^{(0)}_{str/em}$ and ${\cal L}^{(1)}_{\slashT}$, respectively.
For simplicity only one of two possible orderings is shown here.}
\label{edfig1}
\end{figure}

When we consider the sizes of specific contributions to the neutron EDFF, 
the first tree-level contribution comes from the vertex generated by 
the Lagrangian of Eq. (\ref{LTviol2}).  
The lowest-order one-loop graphs are built out of one vertex from 
the $T$-violating Lagrangian (\ref{LTviol-1}) 
and all other vertices from the strong Lagrangian (\ref{Lstr0}). 
The one-loop diagrams that \textit{a priori} could contribute to the EDFF 
are shown in Fig. \ref{edfig1}.  
These one-loop diagrams contribute  
${\cal O}(e\bar{g}_0 f_\pi Q/ (4\pi f_\pi)^2)$ to $J^\mu_{ed}(q)$,
and thus give rise to an EDFF of
the same order as the short-range, tree-level contribution to the EDM,
${\cal O}(em_\star \bar{\theta}/M_{QCD}^2)$.

Because of the isospin structure, the contribution
from Fig. \ref{edfig1}(c) vanishes.
We evaluate diagrams in Fig. \ref{edfig1}
at $v \cdot q =0$, as a consequence of the static nature of the nucleon
in leading order.
Diagram \ref{edfig1}(b) then also vanishes.
Diagram \ref{edfig1}(a)  
gives a non-zero contribution only to the isovector component
of the nucleon EDFF.  

The contributions from Fig. \ref{edfig1} contain 
divergent pieces that can be absorbed by redefining the tree-level 
contribution to the EDFF.
Taking the tree and one-loop contributions, we have
\begin{eqnarray}
d_0 &=& \tilde{d}_0 \\
d_1 &=& \tilde{d}_1 + \frac{eg_{A}\bar{g}_{0}}{8\pi^{2}f_{\pi}}
        \bigg[\bar{\Delta}+2\ln\frac{\mu}{m_{\pi}}\bigg],
\label{d1}
\end{eqnarray}
where $\mu$ is the renormalization scale introduced by 
dimensional regularization, and
$\bar{\Delta}\equiv 2/\varepsilon-\gamma_{E}+\ln 4\pi$  
($\varepsilon=4-d$, with $d$ the spacetime dimension).
Using the Goldberger-Treiman relation  and setting $\mu$ to $m_{N}$,
the piece in Eq. (\ref{d1}) that is non-analytic in $m_\pi^2$
can be written as 
$(eg_{\pi NN}\bar{g}_{0}/4\pi^{2}m_{N})\ln(m_{N}/m_{\pi})$, in agreement 
with the result of Crewther \textit{et al.} \cite{CDVW79}.
The short- and long-range contributions 
are in general of the same size, but a cancellation
is unlikely due to the $\ln m_\pi$ dependence of the pion contribution.
Note that short- and long-range physics cannot
be separated with a measurement of the neutron EDM 
$d_n= d_0-d_1$ alone.
 
\begin{figure}[t]
\begin{center}
\epsfxsize=13cm
\centerline
{\epsffile{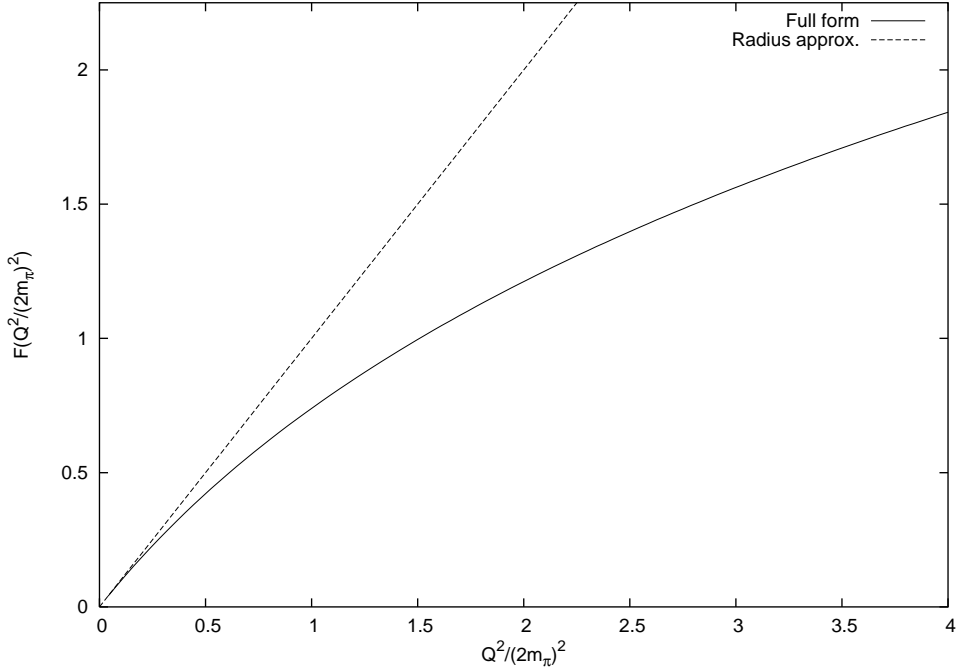}}
\end{center}
\vspace{-0.5cm}
\caption{The function $F(Q^2/(2m_\pi)^2)$
that enters the
isovector electric dipole form factor $F^{(1)}_D$:
$\chi$PT in leading order, Eq. (\ref{F}) (solid line);
and quadratic approximation, Eq. (\ref{radapp}) (dotted line).}
\label{edfig2}
\end{figure}

Because the form factor is given by lowest-order loop graphs,
it depends on the combination $Q^2/m_\pi^2$ only.
Because the only non-vanishing contribution to this order comes
from Fig. \ref{edfig1}(a), the combination is actually
$Q^2/(2m_\pi)^2$.
The EDFF defined in Eq. (\ref{cur}) is found to be given by
\begin{eqnarray}
F_{D}^{(0)}(Q^{2})&=&d_0\\
F_{D}^{(1)}(Q^{2})&=&d_1
              -\frac{eg_{A}\bar{g}_{0}}{12\pi^{2}f_{\pi}}
               \; F\left(\frac{Q^{2}}{(2m_{\pi})^{2}}\right),
\label{FD}
\end{eqnarray}
where\begin{equation}
F(x)=3\left\{\frac{1}{2}\sqrt{1+\frac{1}{x}} \;
           \ln{\left(\frac{\sqrt{1+1/x}+1}{\sqrt{1+1/x}-1}\right)-1}\right\}.
\label{F}
\end{equation}
Note that $F(0)=0$, so indeed this does meet the standard of being 
a form factor.  
The function $F$ is a testable prediction of $\chi$PT:
it is plotted as a function of $Q^2$ in Fig. \ref{edfig2}.

For $x\ll 1$ we can expand $F(x)$ in powers of $x$,
\begin{equation}
F(x) =
       x+{\cal O}(x^2).
\label{radapp}
\end{equation}
This approximation is compared to the full functional form (\ref{F})
in Fig. \ref{edfig2}.
The variation of the form factor with $Q$ can be characterized 
at very small momenta by 
the isovector EDM square radius 
(defined in analogy to the charge square radius), 
\begin{equation}
\langle r_{ed}^{2}\rangle^{(1)}=
-6\left(\frac{dF_{D}^{(1)}}{dQ^{2}}\right)_{Q^{2}=0}
=\frac{eg_{A}\bar{g}_{0}}{8\pi^{2}f_{\pi}}\frac{1}{m_{\pi}^{2}}.
\label{radius}
\end{equation}
While the EDM vanishes in the chiral limit, the radius of the EDFF
is finite.
As was noted in Ref. \cite{scott}, 
$S'= \langle r_{ed}^{2}\rangle^{(1)}/6
=eg_{A}\bar{g}_{0}/48\pi^{2}f_{\pi}m_{\pi}^{2}$ ($-S'$)
is the leading
electromagnetic contribution to the Schiff moment of the 
the proton (neutron).

The mass scale governing the variation of $F(Q^2/(2m_\pi^2))$ 
is in fact $2m_\pi$, since it is generated by the pion cloud.
The relative importance of the $Q^2$ variation  of $F_{D}^{(1)}(Q^2)$
depends in addition on the ratio 
$eg_{A}\bar{g}_{0}/12\pi^{2}f_{\pi} d_1$.
This ratio is a constant in the chiral limit. 
Since we expect it to be ${\cal O}(1)$, 
it is likely that $m_\pi$ sets the scale 
of the variation of $F_{D}^{(1)}(Q^2)$ as well.
Unlike other nucleon form factors, a dipole approximation with a mass
scale close to the rho-meson mass, $m_\rho$, should not be good
even as a
numerical approximation for the EDFF.

All other contributions to the EDFF are formally of
higher order. Next-order effects come from diagrams with one insertion of
an operator with one more derivative.
First, there are operators in 
\begin{eqnarray}
{\cal L}^{(1)}_{str/em}=
\frac{1}{2m_N}\left\{ \bar{N}((v\cdot D)^2- D^2) N
+\frac{g_A}{f_\pi}\left[i\bar{N} (v\cdot D\boldtau\cdot\boldpi) S\cdot D N 
                       + {\rm H.c.} \right]
\right.\nonumber \\
\left.+ \frac{e}{2}\varepsilon_{\mu\nu\rho\sigma}
\bar{N} \left(1+\kappa_0+(1+\kappa_1) \tau_3\right) 
        v^\mu S^\nu N F^{\rho\sigma}+ \ldots\right\}
+ \ldots
\label{Lstr1}
\end{eqnarray}
These are nucleon recoil corrections to the static limit
and photon coupling to the nucleon magnetic moment
(including the isoscalar and isovector anomalous magnetic moments,
$\kappa_0=-0.12$ and $\kappa_1=3.7$ respectively).
Second, there is a new operator in
\begin{equation}
{\cal L}^{(2)}_{\slashT}=\frac{2\bar{h}_{0}}{D}
          \; \boldpi\cdot D_\mu\boldpi \; \bar{N} S^\mu N,
\label{LTviol0}
\end{equation} 
\noindent 
with an undetermined coefficient 
$\bar{h}_{0}={\cal O}(m_\star \bar{\theta}/ f_\pi^2 M_{QCD})
={\cal O}(m_\pi^2 \bar{\theta}/ f_\pi^2 M_{QCD}^2)$.
The corresponding contributions  to the nucleon EDFF
are also calculable, as they are still given by long-range physics
associated with the pion cloud.
Although the strong-interaction corrections depend on  known parameters,
Eq. (\ref{LTviol0}) introduces a new
$T$-violating parameter.
Both types of contributions ---from Eqs. (\ref{Lstr1}) and (\ref{LTviol0})--- 
are suppressed 
by ${\cal O}(Q/M_{QCD})$ relative to the contributions calculated
here.

Note that if ${\cal L}_{\slashT}$ contained other
non-derivative, $T$-violating pion-nucleon interactions,
$\pi_3 \bar{N}N $ and
$\pi_3 \bar{N}\tau_3N$ \cite{barton},
our previous results would remain unchanged.
However, the chiral-symmetry transformation properties
of the $\bar\theta$ term do not yield such terms.

In conclusion,
we have presented results for the electric dipole form factor of the nucleon 
due to the $\bar\theta$ term
in leading order in $\chi$PT.
We have shown that the the EDFF is isovector, with a 
$Q^2$ dependence determined by 
a non-derivative $T$-violating pion-nucleon coupling
and the pion mass,
see Eqs. (\ref{FD}) and (\ref{F}), and Fig. \ref{edfig2}.
Under the assumption that higher-order results are not afflicted by
anomalously-large dimensionless factors, 
the error of our results
at momentum $Q$ should be $\sim Q/m_\rho$.

\vspace{1cm}
\noindent
{\large\bf Acknowledgments}

\noindent
We thank Steve Puglia for discussions
and the Institute for Nuclear Theory at the University
of Washington for hospitality at the beginning of this work.
This research was supported in part by the 
US Department of Energy (WHH, UvK)
and the Alfred P. Sloan Foundation (UvK).

\end{document}